\documentclass[reprint,preprintnumbers,nofootinbib,amsmath,amssymb,aps,floatfix]{revtex4-1}
\usepackage{geometry}

\usepackage{amsmath}  
\usepackage{amsfonts} 
\usepackage{graphicx} 
\usepackage{xcolor}

\newcommand{\ord}[1]{\mathcal{O}(#1)}

\newcommand{\eff}{\mathrm{eff}}

\newcommand{\st}[1]{\mathrm{s}_{#1\Theta}}
\newcommand{\ct}[1]{\mathrm{c}_{#1\Theta}}

\begin{document}
\allowdisplaybreaks

\title{The Inverted Pendulum as a Classical Analog of the 
EFT Paradigm}

\author{Martin Beneke}
\author{Matthias K\"onig}\email{matthias.koenig@uni-mainz.de}
\author{Martin Link}

\affiliation{Physik Department T31, James-Franck-Stra\ss e 1, Technische Universit\"at M\"unchen, D-85748 Garching, Germany}

\date{August 26, 2023}

\begin{abstract}
\noindent 
The inverted pendulum is a mechanical system with a rapidly oscillating pivot point. Using techniques similar in spirit to the methodology of effective field theories, we derive an effective Lagrangian that allows for the systematic computation of corrections to the so-called Kapitza equation. 
The derivation of the effective potential of the system requires non-trivial matching conditions, which need to be determined order by order in the power-counting of the problem. 
The convergence behavior of the series is investigated 
on the basis of high-order results obtained by this 
method. The results from this analysis can be used to determine the regions of parameter space, in which the inverted position of the pendulum is stable or unstable to high precision.\\[0.2cm] 
TUM-HEP-1471/23 
\end{abstract}

\maketitle

\section{The Inverted Pendulum}

\begin{figure}[b]
 \begin{center}
 \includegraphics[scale=1.2]{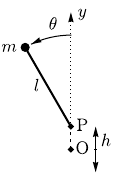}
\end{center}
\vskip-0.4cm
 \caption{Depiction of the inverted pendulum with shaft of length $l$ and point mass $m$. The point O denotes the coordinate origin $y=0$ around which the pivot P oscillates with an amplitude $h$ and a frequency $\omega$. }
 \label{fig:pendulum}
\end{figure}

\noindent
The Kapitza or inverted pendulum is a mechanical system consisting of an inverted rigid pendulum with a rapidly oscillating pivot point P as depicted in Fig.~\ref{fig:pendulum}. 
The system is an interesting exercise for students of classical mechanics due to a curious counter-intuitive phenomenon that occurs for a certain range of parameters: If the frequency of the pivot oscillation is much larger than the natural frequency of the unperturbed system and the amplitude of said oscillation $h$ is smaller (within a certain range) than the length of the pendulum $l$, the system exhibits a stable equilibrium in the upright position \cite{Stephenson_1908, kapitza1951dynamic, LandauLifshitz, Butikov2001}. 

We parameterize the pendulum by its length $l$, the amplitude $h$ and frequency $\omega$ of the driving vertical oscillation of the pivot point, and the angle $\theta$ of the pendulum with respect to the upright position. Using $\theta$ as a generalized coordinate, the potential $V_\mathrm{grav}= m g y$ 
where $g=9.81 \,\mbox{m}/\mbox{s}^2$, 
and the driving oscillation parameterized by $\vec P(t) = (0,h\sin\omega t)$, the Lagrangian of the system can be written as
\begin{eqnarray}
 	\mathbf L &=& m l^2 \left(\frac{1}{2} \dot{\theta}^2 - \frac{g}{l} \cos\left(\theta\right) - \frac{h}{l} \dot{\theta}\omega \sin\left(\theta\right)\cos\left(\omega t\right)\right) 
\nonumber\\
 	&\equiv& ml^2\, L(\theta,\dot\theta,t).
\end{eqnarray}
In the second line and throughout the rest of this work 
we rescale quantities with dimension of energy (e.g.~Lagrangians, potentials, energies) by $ml^2$ for notational 
convenience.
The factor $\cos\omega t$ encodes the external driving oscillation from the pivot point of the pendulum. The range of these parameters is chosen such that the driving oscillation is a fast vibration of the pivot with small amplitude, such that $\sqrt{g/l} / \omega \ll 1$ and $h/l\ll 1$. We then rewrite the Lagrangian such that these ratios are described by a single small parameter $\lambda$, given by the ratio between the natural frequency of the unperturbed pendulum $\omega_0=\sqrt{g/l}$ and the driving frequency $\omega$. The ratio $h/l$ is written as the same small parameter $\lambda$ times an $\mathcal{O}(1)$ number $\rho$, i.e.
\begin{equation}
 \lambda = \frac{1}{\omega}\sqrt{\frac{g}{l}}\,, \qquad
 \frac{h}{l} =  \lambda\rho\,.
\end{equation}
With these definitions, the Lagrangian can be written as
\begin{align}
 L = \frac{1}{2}\dot\theta^2 -\lambda^2\omega^2\cos(\theta)-\lambda\omega\rho \dot\theta \sin\theta\,\cos\omega t \,, \label{eq:lagr_full}
\end{align}
from which the equation of motion can easily be derived:
\begin{align}
 \ddot\theta = \lambda\omega^2 \left(\lambda-\rho\sin\omega t \right) \sin\theta\,. \label{eq:eom_full}
\end{align}

\begin{figure}
 \centering
 \includegraphics[width=0.45\textwidth]{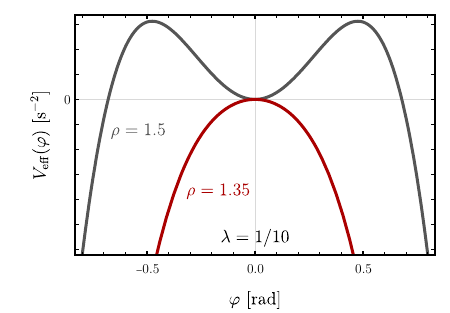}
 \caption{Effective potential from eq.~\eqref{eq:Veff0} for a stable configuration $\rho > \sqrt{2}$ (gray line), and an unstable configuration $\rho < \sqrt{2}$ (red line).}
 \label{fig:Veff0}
\end{figure}

We now want to study the motion of the pendulum across time scales larger than the period of the fast oscillation, imagining a situation where the fast oscillations are too rapid to be resolved by the measurement device and only a time-averaged motion is seen \cite{kapitza1951dynamic,Stephenson_1908,LandauLifshitz}. To this end, one separates the generalized coordinate into a slowly- and a rapidly-oscillating component $\varphi$ and $\eta$, respectively,
\begin{align}
 \theta(t) = \varphi(t)+\eta(t)\,, \label{eq:intro_modesplitting}
\end{align}
where $\varphi(t)$ is assumed to contain frequencies of the order $\mathcal{O}(\omega_0)$ and thus only varies slowly over time. It follows that derivatives acting on the low-frequency mode $\varphi(t)$ are proportional to the frequency $\omega_0=\lambda\omega$ and thus of $\mathcal{O}(\lambda)$:
\begin{align}
 \frac{d^n}{dt^n}\varphi(t) \sim \mathcal{O}(\omega^n\lambda^n) \varphi(t)\,. \label{eq:deriv_scaling}
\end{align}
On the other hand, the rapid-oscillation mode $\eta(t)$ dominantly contains frequencies of order $\mathcal{O}(\omega)$. Since the only source of rapid oscillations is the vibrating pivot, and its amplitude is of $\mathcal{O}(\lambda)$, the fast oscillating component $\eta(t)$ should itself be counted as $\mathcal{O}(\lambda)$. Inserting this mode-splitting into the equation of motion~\eqref{eq:eom_full}, it becomes
\begin{align}
 0 = \ddot\varphi+\ddot\eta -\lambda\omega^2\left( \lambda-\rho\sin\omega t\right) \sin(\varphi+\eta)\,.
\end{align}
Using the $\lambda$ power-counting explained above, we expand this expression in $\lambda$, yielding 
\begin{eqnarray}
0 &=& \ddot\eta + \lambda\rho\omega^2\sin\varphi\sin\omega t \nonumber \\
&& +\,\ddot\varphi - \lambda\omega^2\left( \lambda\sin\varphi - \rho\, \eta\, \cos\varphi \, \sin\omega t\right) \nonumber\\  
&& +\, \ord{\lambda^3}\,,
\end{eqnarray}
where the first line of the right-hand side is of order $\ord{\lambda}$ and the second of order $\ord{\lambda^2}$. Because the equation has to be satisfied at each order in $\lambda$ separately, it splits into two equations:
\begin{align}
\begin{aligned}
 \ddot\eta &= - \lambda \rho \omega^2 \sin\varphi \sin\omega t  \,, \\
 \ddot\varphi &= \lambda\omega^2\left( \lambda\sin\varphi - \rho\, \eta\, \cos\varphi \, \sin\omega t\right)  \,. \label{eq:eom_split_lp} 
\end{aligned}
\end{align}
In the equation for $\eta$, the slowly varying function $\varphi$ can be regarded as constant
over the time scale of one oscillation of $\eta$.
The equation can thus be integrated directly in $t$, yielding
\begin{align}
 \eta = \lambda\rho \sin\varphi\,\sin\omega t\,. 
\label{eq:lp_eta_solution}
\end{align}
The integration constant must be set to zero, since it 
would constitute a slowly varying component that is absent 
from $\eta$ by construction. Inserting the solution for 
$\eta$  into the second equation of~\eqref{eq:eom_split_lp} one finds
\begin{align}
 \ddot\varphi = \lambda^2\omega^2\left(1-\rho^2\cos\varphi\,\sin^2\omega t \right) \sin\varphi\,.\label{eq:slowsolutionLP}
\end{align}
Since the function $\varphi(t)$ varies only slowly with time compared to the driving oscillation, we rewrite $\sin^2\omega t = \frac{1}{2}(1-\cos 2\omega t)$ and keep only the constant term. This can be interpreted as averaging the equation over one period of the driving oscillation. The equation of motion becomes
\begin{eqnarray}
 \ddot\varphi &=& \lambda^2 \omega^2\left(1-\frac{\rho^2}{2}\cos\varphi \right) \sin\varphi
\nonumber\\
&\equiv& -\frac{d}{d\varphi}V_\eff(\varphi)\,,
\label{eq:averg_phi}
\end{eqnarray}
which is known as the Kapitza equation. The motion of the pendulum, when averaged over the rapid 
oscillations, can be described as the motion in a time-independent {\em effective} potential, which by 
integrating the right-hand side of the first line is found to be
\begin{equation}
 V_\eff(\varphi) = \lambda^2\omega^2 \left(\cos\varphi- 1 + \frac{\rho^2}{4}\sin^2\varphi\right). 
\label{eq:Veff0}
\end{equation}
The integration constant is chosen such that $V_\eff(0) = 0$. The effective potential has a stable equilibrium in the upright position $\varphi = 0$ when $\rho > \sqrt{2}$, which follows from 
requiring the second derivative at the origin to 
be positive, $V_\eff''(\varphi) > 0$. The potential is shown in Fig.~\ref{fig:Veff0} for a stable ($\rho>\sqrt{2}$) and an unstable ($\rho < \sqrt{2}$) configuration. 

\begin{figure}
 \centering
 \includegraphics[width=0.45\textwidth]{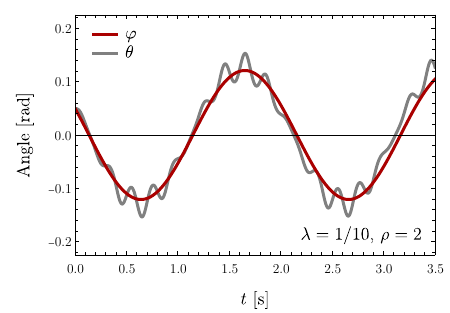}
 \caption{Numerical solutions of the exact system (gray line) and the leading-order effective system (red line) for the parameters $\lambda=1/10$ and $\rho =2$.}
 \label{fig:traj_lp}
\end{figure}

From the above results, one can derive the averaged pendulum trajectory $\varphi(t)$, shown for the parameters $\lambda=1/10$ and $\rho=2$ in Fig.~\ref{fig:traj_lp} as the red line. The figure also shows the numerical solution of the full system 
\eqref{eq:eom_full}, in which the rapidly oscillating components can still be seen. One notices that the trajectory of $\varphi$ approximates the low-frequency component of $\theta$ well. But after few periods of the slow oscillation, the trajectories begin to diverge. To obtain a more precise approximation, corrections beyond the leading order in $\lambda$ need to be included. We remark that to compare the two trajectories, one needs to account for the fact that the initial conditions on $\theta$ in the full system and on $\varphi$ in the effective system are not identical and need to be matched, as will be discussed in Sec.~\ref{sec:initconds} in detail.

It is interesting to note that the effective Lagrangian is that of a closed system, since it does not explicitly depend on the time $t$, which is not the case for the full system. This can be understood from the fact that in the full system, the time-dependence of the Lagrangian arises through the rapid oscillations: The vibrating pivot periodically increases and decreases the energy of the system over the time scale of the fast oscillation. Over much longer time scales, this effect averages out and the system becomes a closed one. 

The above discussion shows that high-frequency fluctuations that may not be resolved by observation, can nevertheless change qualitatively the low-frequency oscillatory behavior, here by creating a stable equilibrium in the upright position. A similar phenomenon arises in quantum field theory, where short-distance quantum fluctuations can change the ground-state / vacuum of the system~\cite{Coleman:1973jx}. This is currently of great interest since a competition between the quantum fluctuations of the top quark and those of the electroweak and Higgs boson determines whether the ground-state in the effective Higgs potential is stable or metastable~\cite{Buttazzo:2013uya}.

The derivation of the effective potential \eqref{eq:Veff0} presented above is unsatisfactory for its intransparency as to how it can be systematically generalized to higher orders in the approximations employed. In both equations of motion~\eqref{eq:eom_split_lp} we have made approximations motivated by heuristic scaling arguments: In the equation for the rapid mode $\eta$, the slow mode $\varphi$ has been treated as static, supported by the scaling argument~\eqref{eq:deriv_scaling}. At higher orders in $\lambda$, derivatives of $\varphi$ should contribute but their treatment is unclear. The same issue arises with the averaging procedure leading to~\eqref{eq:averg_phi}. 

Previous attempts to solve these issues focused on the improvement of the time-averaging approach \cite{Rahav_2003,Grundy_2019,Maggia_2020} often using multi-scale methods \cite{Kuehn2015}, and mathematical analysis of the exact equations of motion \cite{Sanz-Serna_2009}. Furthermore, many authors have discussed more accurate approximations of the stability boundaries \cite{Butikov_2011, ButikovStability2018,blackburn1992stability,MOND199377,insperger2002stability,insperger2003stability}, but to our knowledge a method to obtain the systematic expansion of the effective potential in $\lambda$ has not been discussed before.

In order to address this problem, we draw inspiration from the methodology of \emph{effective (quantum) field theories}. Throughout all areas of physics, an important concept in approaching complex systems is to identify the degrees of freedom that are of subleading importance for the observable of study. For example, the motion of an electrically charged object in the field generated by the distribution of charged point particles contained in a finite volume can be described by the multipole moments of the distribution when the object is far away compared to the size of the system \cite{jackson2021classical}.  The calculation of the multipole moments {\em does} require knowledge of the ``microscopic'' distribution of the point particles, but the details are not relevant for the object moving on ``macroscopic'' scales.

The association of effects and their origins with characteristic energy or length scales is key to the concept of effective field theory. One then establishes a parametric limit in which the expressions describing the system can be expanded. The resulting effective theory is then a simpler albeit approximate description of the system, in which the subleading effects can in principle be included as power-corrections in an expansion around the parametric limit to improve the approximation to the desired accuracy.

Effective theories play a prominent role in quantum field theories, where quantum fluctuations are always present on all scales. To focus on the scales {\em directly} probed by observations of the physical system, one removes (``integrates out'') the shorter-ranged quantum fluctuations to arrive at an effective description in terms of the relevant degrees of freedom~\cite{Weinberg:1978kz,Polchinski:1992ed,manohar2018EFT}. The inverted pendulum with $\omega\gg \omega_0$ provides a classical analogy of removing a small time scale, provided by the inverse of the rapid oscillation frequency of the pivot point. The condition $h\ll l$ implies that the rapid oscillations are ``weak'', so that they can be treated perturbatively in the small parameter $\lambda=\omega_0/\omega$.
 
An important aspect in finding effective descriptions is the ability to go beyond the leading approximations in a systematic fashion by including higher-order terms in the power-expansion in $\lambda$. While simple from a conceptual point of view, the treatment of subleading-power effects often brings technical subtleties not present at the leading power of the expansion.

In the following, we provide a derivation of the effective Lagrangian for the inverted pendulum that differs from the approach demonstrated above, is close in spirit to the mode decomposition common to effective quantum field theories, and can be systematically extended to any order in the expansion parameter $\lambda$. We first show the  derivation of the first non-trivial power correction in detail and then give results based on a higher-power analysis.


\section{Subleading-Power Analysis}

\subsection{Mode Expansion}

\noindent We now present a more systematic approach to the problem. Our ansatz is to make the frequency-separation manifest by factoring out all fast oscillations, which come in multiples of the large frequency $\omega$, from the generalized coordinate in terms of the modulated Fourier expansion~\cite{Sanz-Serna_2009}
\begin{align}
 \theta(t) = \sum_{n=0}^\infty \left[ \Psi_n(t)\sin(n\omega t) + \Phi_n(t) \cos (n\omega t)\right]\,. \label{eq:mode_decomp1}
\end{align}
By construction, the coefficient mode functions $\Psi_n$ and $\Phi_n$ of the rapidly varying trigonometric functions only vary slowly with the time $t$, specifically with time scales of the order $\mathcal O(1/\omega_0)$. The low-frequency mode $\varphi$ from the previous section is then the contribution from $n=0$. Including only discrete values of $n$ is justified by the fact that the sole origin of high-frequency oscillations is the sinusoidal source term proportional to $\lambda\cos\omega t$ in the Lagrangian~\eqref{eq:lagr_full}. 
We now note that the source term coupling the coordinate $\theta$ to the high-frequency external vibration comes with one power of $\lambda$, implying that 
the amplitude of $n$-th harmonic of $\omega$ is naturally suppressed by $\lambda^n$. We also deduce that the even-$n$ terms can only involve $\Phi_n$ whereas the odd-$n$ terms only have $\Psi_n$ coefficients. To solve the equations for 
$\Phi_n,\Psi_n$ successively in powers of $\lambda$, we  
expand the mode functions for $n > 0$ in the power-counting parameter, 
\begin{align}
 \Psi_n(t) &= \sum_{m=0}^\infty \psi_{nm}(t)\,, &
 \Phi_n(t) &= \sum_{m=0}^\infty \varphi_{nm}(t)\,,
\label{eq:modelambdaexp}
\end{align}
and identify the zero mode (recall $\Psi_0 = 0$)as 
\begin{equation}
\Theta(t) \equiv \Phi_0(t) \,.
\label{eq:zeromodedef}
\end{equation} 
In these series, the $m$-th terms scale as    
\begin{equation}
 \psi_{nm}(t),\, \varphi_{nm}(t) \ \sim \ \lambda^{n+m}
\,.\label{eq:mode_exp0}
\end{equation}
Since by assumption these functions are slowly varying, their time derivative is accompanied by a factor of $\omega_0 = \lambda\omega$, hence their $k$-th derivative counts as
\begin{align}
 \psi_{nm}^{(k)}(t),\, \varphi_{nm}^{(k)}(t) \ \sim \ \lambda^{n+m+k}\,.\label{eq:mode_exp1}
\end{align}
For later convenience, we introduce the abbreviations
\begin{eqnarray}
\label{eq:trigconv}
&& \sin (n\cdot\Theta(t)) = \st{n}\,,\;\;
\cos (n\cdot\Theta(t)) = \ct{n} \,.\quad
\end{eqnarray}
We emphasize that unlike the fast components of $\theta(t)$, the 
zero mode $\Theta(t)$ and consequently $\st{n}$, $\ct{n}$ are  
not expanded in $\lambda$ in the following.
 
We now insert eq.~\eqref{eq:mode_decomp1} into the Lagrangian~\eqref{eq:lagr_full} and expand to the desired order in $\lambda$. Since all factors of $\sin\omega t$ and $\cos\omega t$ are explicit in the mode expansion of the coordinate $\theta$, the Lagrangian itself automatically decomposes into Fourier modes,  casting it into the form:
\begin{eqnarray}
 L &=& \sum_{k=0}^\infty \left[ A_k(\Phi,\dot\Phi,\Psi,\dot\Psi)\cdot\cos(k\omega t) \right.
\nonumber \\ 
&& \left.\hspace*{0.6cm} + \,B_k(\Phi,\dot\Phi,\Psi,\dot\Psi)\cdot\sin(k\omega t) \right]. 
\label{eq:lagr_modes}
\end{eqnarray}
The coefficients of the rapid oscillations $\cos (k\omega t)$ and $\sin (k\omega t)$ in this equation are not to be confused with the ones in the mode decomposition of the generalized coordinate $\theta$ appearing in the decomposition~\eqref{eq:mode_decomp1}, as the full Lagrangian~\eqref{eq:lagr_full} is non-linear in $\theta$. Products of different fast Fourier modes of $\theta$ then decompose into a sum of terms in the Lagrangian \eqref{eq:lagr_modes} after making use of standard trigonometric identities.

The Lagrangian~\eqref{eq:lagr_modes} is still equivalent to the one of the full system, and describes its dynamics at all frequency scales. The slowly oscillating behavior of the system, which we are ultimately interested in, is given by the $k=0$ term in the sum, $L_{k=0} \equiv A_0$. This expression contains contributions from the $n=0$ modes of $\theta$ as well as terms in which the higher modes combine to yield a slowly-varying function. This is equivalent to the averaging procedure in the previous section, where we expanded the solution of the low-frequency mode~\eqref{eq:slowsolutionLP} using trigonometric identities and kept only the term without $\cos(k \omega t)$. To determine the mode functions $\psi_{nm}(t)$, $\varphi_{nm}(t)$, we insert the mode expansion into the equation of motion of the full system, and solve it separately mode by mode and order by order in the expansion in $\lambda$. 

We start at the equation of motion \eqref{eq:eom_full} of the full system, 
\begin{align}
 \ddot\theta = \lambda^2\omega^2\sin\theta - \lambda\rho\omega^2\sin\theta \sin\omega t\,,
\end{align}
and insert eqs.~\eqref{eq:mode_decomp1} and~\eqref{eq:modelambdaexp} into the above expression. 
We then expand in powers of $\lambda$ according to the defined scaling rules~\eqref{eq:mode_exp0}, \eqref{eq:mode_exp1} for all the objects. For example, up to order $\lambda^2$, the equation of motion becomes:
\begin{eqnarray}
 0&=&\lambda\omega^2\sin(\omega t) \left[\rho \st{{}}- \frac{\psi_{10}}{\lambda} \right] \nonumber \\
 &&+\,\lambda^2 \left(\left[\frac{\ddot\Theta}{\lambda^2} -\omega^2\st{{}} + \frac{\rho^2\omega^2}{2}\ct{{}}\frac{\psi_{10}}{\lambda}\right]\right. \nonumber \\
 && + \left.\omega\cos(\omega t) \left[ \frac{2\dot\psi_{10}}{\lambda^2}-\omega\frac{\varphi_{11}}{\lambda^2} \right] \right.\label{eq:full_eom_expanded}
 \\
 &&-\left. \frac{\omega^2}{2}\cos(2\omega t) \left[ \rho\ct{{}} \frac{\psi_{10}}{\lambda}+ \frac{8\varphi_{20}}{\lambda^2} \right]
 \right)+\ord{\lambda^3}\,.\nonumber 
\end{eqnarray}
The equation needs to be fulfilled at each order in $\lambda$ separately, as well as for each harmonic of $\omega$. In eq.~\eqref{eq:full_eom_expanded} this means that each square bracket has to vanish independently, defining a system of equations. One then solves them at each order in $\lambda$, inserting the solutions in the equations at the next higher order until all mode functions $\psi_{nm}$, $\varphi_{nm}$ are determined. In the above example, this means we start by setting the expression in square brackets in the first line to zero, yielding
\begin{align}
 \psi_{10}=\lambda\rho\st{{}}\,.
\end{align}
Inserting this solution into the $\ord{\lambda^2}$ terms of eq.~\eqref{eq:full_eom_expanded} and requiring each bracket to vanish, one finds the solution for the next-to-leading order modes:
\begin{align}
 \varphi_{20}(t) &= -\frac{\lambda^2\rho^2}{8}\ct{{}}\st{{}}\,, &
 \varphi_{11}(t) &= \frac{2\lambda\rho}{\omega} \dot\Theta\ct{{}}\,.
\end{align}
The square bracket in the second line of eq.~\eqref{eq:full_eom_expanded} contains no further information about any $\varphi_{nm}$, $\psi_{nm}$ and can be discarded -- it contains instead the leading-power equation of motion of the zero-mode $\Theta(t)$. Using the above results, we can insert them into $L_{k=0}$ to obtain the \emph{effective Lagrangian} up to order $\ord{\lambda^4}$:
\begin{widetext}
\begin{align}
\begin{aligned}
  L_\eff =  &\frac{1}{2}\dot\Theta^2 \left( 1- \frac{3}{2}\lambda^2\rho^2\ct{{}}^2\right)
  -\lambda^2\omega^2\left( \ct{{}} - \frac{\rho^2}{8}\ct2 \right)
 + \frac{\lambda^4\omega^2\rho^2}{16}\left( \ct{{}}- \frac{\rho^2}{2}\ct2
 -\ct3 + \frac{5}{32}\rho^2\ct4
 \right)\\
 &+\lambda^2\rho^2\st{{}} \ddot{\Theta} \left(\frac{5}{64}\lambda^2\rho^2\ct{3}-\lambda^2\ct{{}}^2- \frac{5}{\omega^2}\ct{{}}\dot\Theta^2 \right) + \frac{\lambda^2\rho^2}{4\omega^2}\ct{{}}^2\ddot\Theta^2 + \frac{3\lambda^2 \rho^2 }{2\omega^2}\ct{{}}^2\dot\Theta\Theta^{(3)} + \ldots
\label{eq:lageps4}
\end{aligned}
\end{align}
\end{widetext}
The second line adds a subset of $\mathcal{O}(\lambda^6)$ 
terms to be discussed below. 
This effective Lagrangian determines the dynamics of the system when the fast oscillations at the scale $\omega$ are not resolved by the measurement. By determining the equation of motion for $\Theta$ and solving it, we obtain the motion of the pendulum when the fast fluctuations are averaged over. Some comments are in order. 

First, the above procedure can be generalized to an arbitrary order in $\lambda$, and at any order one finds the equations determining the mode functions to be purely algebraic and never differential. This is due to the fact that derivatives of mode functions are always power-suppressed. Thus, equations involving derivatives of a given mode function appear at higher orders in $\lambda$ than equations involving the same function without derivatives, but the corresponding mode is always already determined at lower orders through an algebraic equation. 

Second, the effective Lagrangian does not explicitly 
depend on time, and the averaged system remains conservative 
to any order in the $\lambda$ expansion. This is not true for the full system, as the pivot oscillation periodically injects energy to and removes energy from the system. However, 
within the domain of validity of the  expansion in 
$\lambda$, the effect averages out and
rapid oscillations do not transfer energy to the 
pendulum. The system is effectively closed.

Third, at  $\ord{\lambda^4}$ the effective Lagrangian 
contains a coordinate-dependent kinetic term of the 
form $\dot{\Theta}^2 \ct{{}}^2$, which is a 
general feature of the slow-motion dynamics in rapidly 
oscillating potentials \cite{Rahav_2003, RahavFollowup}. 
Furthermore, starting at order $\ord{\lambda^6}$, the effective 
Lagrangian contains higher derivatives of $\Theta$. These 
terms are displayed explicitly in the second line of 
\eqref{eq:lageps4}. In general, at $\ord{\lambda^{2n+2}}$ ($n>0$) any monomial of up to the $(2n-1)$-th derivative of $\Theta$ with a total of $2 n$ derivatives can appear. For example, at order $\ord{\lambda^6}$:
\begin{equation}
\dot{\Theta}^4, \dot{\Theta}^2 \ddot{\Theta},\dot{\Theta} 
\Theta^{(3)},
\ddot{\Theta}^2.
\end{equation} 
Ordinarily, the presence of such terms in the Lagrangian indicates an Ostrogradsky instability of the system, i.e.~that the system's energy is unbounded from below~\cite{Ostrogradsky}. However, in an effective description higher-derivative terms do not lead to an inconsistency if they are power-suppressed. As such they must be treated as perturbations to the leading-power Lagrangian that does not involve higher derivatives. Using power-counting arguments, we will show below that for the inverted pendulum, the higher-derivative terms and higher powers of $\dot{\Theta}^2$ can be traded for subleading terms with fewer derivatives by means of the equations of motion and energy conservation.

The emergence of higher-derivative terms in the systematic treatment of the slow-motion dynamics of the inverted pendulum presents another similarity to the case of effective relativistic quantum field theories, in which the effective Lagrangian can contain kinetic terms with more derivatives than the canonical one, 
$\frac{1}{2}\partial_\mu\phi\partial^\mu\phi$. For example, a four-derivative term $(\partial_\mu\phi\partial^\mu\phi)^2/\Lambda^2$, where $\Lambda$ represents the large momentum scale, would cause an instability and further imply additional tachyonic modes 
violating relativistic causality. However, for slowly 
varying fields compared to the scale $\Lambda$, the 
power-suppression of this term implies that it should be eliminated by perturbative field redefinitions. At leading order in the power-counting of the theory, this procedure is equivalent to using the equations of motion of the field. 

These remarks also clarify a potentially troublesome 
point, namely that the presence of higher-derivatives of $\Theta$ in the effective Lagrangian for the 
inverted pendulum does not imply that the solution to 
the effective equation of motion needs initial conditions 
beyond those for $\Theta$ and $\dot{\Theta}$. 
Instead, the initial conditions follow from matching 
to those of the full system, as detailed below.

\subsection{Effective Potential}

\noindent We now show that the higher derivatives and higher powers 
of $\dot{\Theta}^2$ can be eliminated, and the 
effective Lagrangian converted into the canonical 
form
\begin{align}
 L_\eff = \frac{1}{2}\dot\Theta^2 - 
V_\eff(\Theta, E_{\rm tot})\,
\end{align}
with an effective potential, which beyond the well-known 
leading approximation \eqref{eq:Veff0} turns out to depend 
on the total energy of the system.

Since the effective Lagrangian in the previous subsection 
does not depend explicitly on time, the Noether charge 
associated with the time-translation symmetry defines the 
conserved 
energy. For a Lagrangian depending on higher derivatives 
of the generalized coordinate $\Theta$, 
the relevant expression is~\cite{anderson1973noether}
\begin{widetext}
\begin{align}
 E_{\rm tot}\! = \sum_{k=1}^\infty\sum_{j=0}^{k-1} (-1)^{j}\Theta^{(k-j)}\frac{d^{j}}{dt^{j}}\frac{\partial L_\eff}{\partial\Theta^{(k)}} \!- L_\eff \,, 
\label{eq:noether}
\end{align}
which generalizes the standard expression for the total energy. From the effective Lagrangian \eqref{eq:lageps4}, we 
obtain
\begin{align}
\begin{aligned}
 E_\mathrm{tot} =
 &\frac{1}{2}\dot\Theta^2\left(1- \frac{3\lambda^2\rho^2}{2}\ct{{}}^2 \right) + \lambda^2\omega^2 \left(\ct{{}}- \frac{\rho^2}{8}\ct{2} \right)
 + \frac{\lambda^4\rho^2\omega^2}{16}\left(-\ct{{}}+ \frac{\rho^2}{2}\ct{2}+\ct{3}- \frac{5\rho^2}{32}\ct{4} \right) \\
 &- \frac{5\lambda^2\rho^2}{4\omega^2}\ct{{}}^2\ddot\Theta^2
 + \frac{5\lambda^2\rho^2}{2\omega^2}\ct{{}}^2\dot\Theta\Theta^{(3)}
-\frac{5\lambda^2\rho^2}{\omega^2}\dot\Theta^2\ct{{}}\st{{}}\ddot\Theta+\ldots\,,
\end{aligned}
\label{eq:noethertohigherorder}
\end{align}
\end{widetext}
where in the second line again only the higher derivative 
terms at $\mathcal{O}(\lambda^6)$ are shown. 
We note that the leading-order expression for the 
energy does not depend on higher derivatives and 
higher powers of $\dot{\Theta}^2$. It is therefore 
possible to define an effective potential by 
successively solving
\begin{align}
 E_{\rm tot} = \frac{1}{2}\dot\Theta^2 + V_\eff(\Theta, 
E_{\rm tot}) 
\label{eq:total_energy}
\end{align}
order by order in $\lambda$ to eliminate the higher 
derivatives of $\Theta$ and powers of  $\dot{\Theta}^2$, 
which yields\footnote{
The superscript $(n)$ on $V_{\rm eff}$, $E_{\rm tot}$ 
and $L_{\rm eff}$ here and below denotes the 
$\mathcal{O}(\lambda^n)$ term in the corresponding quantity.}
\begin{align}
V_\eff(\Theta,E_{\rm tot}) \equiv \sum_{n=1}^\infty
V_\eff^{(2n)}(\Theta,E_{\rm tot})\,. 
\label{eq:potential_series}
\end{align}
The effective potential obtained through this procedure 
depends on the total energy of the effective system. 
Expanding $E_{\rm tot}$ and $V_\eff$ in powers of $\lambda$, 
$V_\eff^{(2n)}$ will generally depend on 
$E_\mathrm{tot}^{(2m)}$ with $m<n$. 

Here we demonstrate the procedure up to the next-to-leading 
order in the $\lambda$ expansion. Beginning with the leading 
order, eq.~\eqref{eq:noether} becomes 
\begin{align}
\begin{aligned}
    E_{\rm tot}^{(2)} 
    &= \dot\Theta \frac{\partial L^{(2)}_\eff}{\partial \dot\Theta} - L^{(2)}_\eff\,.
    \end{aligned}
\end{align}
Combining this with eq.~\eqref{eq:total_energy}, one 
obtains
\begin{align}
 V^{(2)}_\eff = \lambda^2\omega^2 \left(\ct{{}} + \frac{\rho^2}{4}\st{{}}^2 \right) + \delta V^{(2)}_\eff\,,
\end{align}
where $\delta V_\eff^{(2)} = -\lambda^2\omega^2$ is a constant shift introduced in order to adjust the potential 
such that $V_\eff(0) = 0$. This order reproduces the 
familiar result \eqref{eq:Veff0}.

At the next order, from the $\mathcal{O}(\lambda^4)$  
terms of \eqref{eq:noethertohigherorder}, 
the procedure yields the following equation to solve 
\begin{eqnarray}
&&V_\eff^{(4)}+\delta V_\eff^{(4)}
\stackrel{!}{=}-\frac{3\lambda^2 \rho^2}{4}
\ct{{}}^2\dot\Theta^2 
\nonumber\\
&&+\, \frac{\lambda^4\rho^2\omega^2}{16}\,\bigg(\!\!-\ct{{}}+ \frac{\rho^2}{2}\ct{2}+\ct{3}- \frac{5\rho^2}{32}\ct{4} \!\bigg).
\;\quad
\end{eqnarray}
The $\dot\Theta^2$ term is eliminated using eq.~\eqref{eq:total_energy} at $\mathcal{O}(\lambda^2)$, expressing it as $\dot{\Theta}^2 = 2 \,(E_{\rm tot}^{(2)} - V_\eff^{(2)})$. 
Tweaking the offset $\delta V_\eff^{(4)}$ such that $V_{\rm eff}(0)=0$, we obtain
\begin{eqnarray}
 V^{(4)}_\eff &=
 &\lambda^4\rho^2\omega^2\left\{
 \frac{17}{16}\ct{{}} - \frac{3}{4}\ct2+ \frac{7}{16}\ct3 \right. \nonumber \\ &&\left. 
 -\, \frac{3}{4} + \rho^2\left(\frac{1}{32}\ct2 - \frac{29}{512}\ct4 + \frac{13}{512} \right)\right\}\nonumber \\
 && +\,\frac{3}{2}\lambda^2\rho^2 \st{{}}^2 E_\mathrm{tot}^{(2)}\,.
\label{eq:effpotsubl}
\end{eqnarray}

When working beyond the fourth order in $\lambda$, 
one has to 
eliminate $\dot{\Theta}^2$ consistently to sufficiently 
high order. Specifically, to obtain the 
$\mathcal{O}(\lambda^{2 n})$ term, $V_\eff^{(2n)}$, 
in the effective potential, one subtracts $\frac{1}{2} 
\dot{\Theta}^2$ from the right-hand side of 
\eqref{eq:noethertohigherorder} and eliminates 
powers of $\dot{\Theta}^2$ in the remaining expression 
by the solution of 
\eqref{eq:total_energy} to order $\lambda^{2 n-2}$, which includes  
the previously determined potential 
terms  $V_\eff^{(2k)}$ with $k<n$.

Starting from $\mathcal{O}(\lambda^6)$, higher derivative terms appear, which can 
be likewise eliminated by taking the appropriate 
number of derivatives of \eqref{eq:total_energy} and 
proceeding as described above for the higher powers of 
$\dot{\Theta}^2$.

The total energy, which first appears at $\mathcal{O}(\lambda^4)$ in the effective potential is determined by the initial conditions at some time $t_i$, thus expressing it through $\Theta(t_i)\equiv \Theta_i$ and $\dot\Theta(t_i)\equiv \dot\Theta_i$. At $\mathcal{O}(\lambda^4)$, only
\begin{align}
 E_{\rm tot}^{(2)} =\frac{1}{2} \dot\Theta^2_i + V_\eff^{(2)}(\Theta_i)
\label{eq:icLO}
\end{align}
is needed. The initial condition of the effective system is related to 
that of the full one as will be explained in the following section. 

With the above results in place, we have an expression for $V_\eff$ that defines a genuine potential, i.e. a function of only $\Theta$ and no derivatives. The result can easily be generalized to an arbitrary order in the expansion in $\lambda$ by iteratively applying the procedure above, order by order. We give these results in Appendix~\ref{app:potential} to 
$\mathcal{O}(\lambda^8)$.

\subsection{Initial-Condition Matching}
\label{sec:initconds}

\noindent When deriving the pendulum motion $\Theta(t)$ from the effective potential and comparing it to the solution of the full system, one has to account for the fact that the initial conditions for the full and the effective system differ, since the full solution still contains the rapid driving oscillations. For example, the full system could have no initial velocity, because at the initial time the slow and fast oscillation accidentally cancel out. The effective system, containing only the slow mode, would then have a non-zero initial velocity. Therefore, when attempting to compare solutions, as shown in Fig.~\ref{fig:traj_lp}, the initial conditions on the full trajectory $\theta(t)$ and the effective one $\Theta(t)$ need to be carefully related to each other. 

As an aside, in applications of effective field theory 
to scattering in quantum field theory, the matching of 
initial conditions is unnecessary, as the asymptotic 
in- and out-states of free particles at $t\to \mp \infty$ 
are expressed in terms of the same degrees of freedom. 
The situation discussed here for the inverted pendulum 
is analogous to the matching of correlation functions of fields, 
when the full-theory and effective theory field variables are 
different. This arises, for example, in the effective 
theory for the long-wavelength fluctuations of a 
massless scalar field on the De Sitter space-time background 
\cite{Cohen:2020php,Cohen:2021fzf}.

We begin by rewriting the mode expansion~\eqref{eq:intro_modesplitting} and solving it for the effective coordinate $\Theta$ at a given initial time $t_i$,
\begin{align}
\begin{aligned}
 \Theta_i &= \theta_i \\
        &- \sum_{n=1}^\infty \left[ \Psi_n(t_i)\sin(n\omega t_i) + \Phi_n(t_i) \cos (n\omega t_i)\right]\,. \label{eq:mode_decomp2}
\end{aligned}
\end{align}
Here we separated the zero mode $\Theta$ from the higher ones and the coordinates with subscript $i$ are meant to be evaluated at the initial time, $\Theta_i = \Theta(t_i)$. Inserting the obtained solutions for the various mode functions $\Psi_n$ and $\Phi_n$ allows one to relate the initial conditions $\theta_i$ to those of the effective coordinate $\Theta_i$. Taking derivatives does the same for the initial angular velocities:
\begin{align}
\begin{aligned}
 \dot\Theta_i &= \dot\theta_i \\
 &- \sum_{n=1}^\infty \Big[ 
 \dot\Psi_n(t_i) \sin (n\omega t_i)+\dot\Phi_n(t_i)\cos(n\omega t_i) \\
 &+ n\omega \big( \Psi_n(t_i)\cos(n\omega t_i) - \Phi_n(t_i)\sin(n\omega t_i)\big) 
 \Big]\,.
\end{aligned}
\end{align}
 At $\mathcal{O}(\lambda)$, the above results translate to
\begin{align}
 \begin{aligned}
    \Theta_i &= \theta_i - \lambda\rho\sin\Theta_i \sin \omega t_i \,, \\
    \dot\Theta_i &= \dot\theta_i - \lambda\rho 
    \omega\sin\Theta_i\cos\omega t_i 
    \,.
 \end{aligned}\label{eq:lp_init_conds}
\end{align}
Note that the second equation does not come with a term involving $\dot\Theta_i$, as $\dot\Theta_i$ is suppressed by an additional power of $\lambda$. Both sides of the equations contain the initial conditions $\Theta_i$, $\dot\Theta_i$ which we aim to solve for. However, they always appear on the right-hand sides in power-suppressed terms. Therefore, the system can be solved iteratively by expanding both $\Theta_i$ and $\dot\Theta_i$ in powers of $\lambda$. At 
$\mathcal{O}(\lambda)$, this amounts to setting
\begin{align}
\begin{aligned}
 \Theta_i &= \theta_i - \lambda \rho \sin\theta_i \sin\omega t_i\,, \\
 \dot\Theta_i &= \dot\theta_i - \lambda\rho \left(\omega\sin\theta_i\cos\omega t_i \right) \,.
\end{aligned}
\end{align}
We give the results to higher order in Appendix~\ref{app:init_conds}. It is important to recall that, at higher orders in power-counting, higher derivatives are introduced again. They can be eliminated with the method described in the last section. With these results, one can derive numerical solutions for the effective system and compare them directly to the ones for the full system, as shown in 
Fig.~\ref{fig:traj_lp} above and Fig.~\ref{fig:traj_slp} below.

The initial conditions are also needed to eliminate the $E_{\rm tot}$ terms in the effective potential, as mentioned in the previous section for the case of $E_{\rm tot}^{(2)}$ in $V_\eff^{(4)}$, where one only needed the leading-power 
expression \eqref{eq:icLO}. In general, at 
$\ord{\lambda^{2n}}$, eq.~\eqref{eq:total_energy} 
evaluated at initial time $t_i$ gives
\begin{align}
 \dot\Theta_i^2 = 2\sum_{j=1}^{n}\left( E_{\rm tot}^{(2j)}-V_\eff^{(2j)}(\Theta_i,E_{\rm tot})\right)\,,
\end{align}
where the argument of $V_\eff^{(2j)}$ involves the expansion terms $E_\mathrm{tot}^{(2m)}$ with $m$ taking values up to $j-1$. Assuming 
that $E_\mathrm{tot}^{(2m)}$ is known up to 
$m=n-1$, the above equation can be solved for 
$E_{\rm tot}^{(2 n)}(\Theta_i,\dot{\Theta}_i)$. 
Finally, $\Theta_i,\dot{\Theta}_i$ are expressed in terms  
of the initial conditions $\theta_i,\dot{\theta}_i$ 
of the full system to order $\lambda^{2n}$ as described 
above to find $E_{\rm tot}^{(2 n)}(\theta_i,
\dot{\theta}_i)$ after re-expanding the entire expression 
in $\lambda$.


\section{Discussion of Results}

\begin{figure}
 \centering
 \includegraphics[width=.45\textwidth]{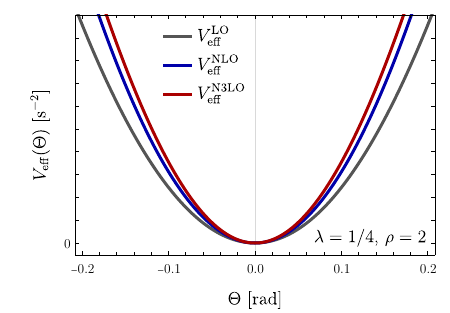}
 \caption{Effective potentials $V_\eff^{NnLO}$, where $n$ denotes the order in power-counting, with $\lambda=1/4$ and $\rho=2$ and initial conditions $\theta(0)=0.05$, $\dot\theta(0)=-0.03\mathrm{s}^{-1}$.}
 \label{fig:Veff3}
\end{figure}

\subsection{Effective potential and 
low-frequency motion}

\noindent We begin by comparing the effective potentials at different orders up to order $\ord{\lambda^8}$. The explicit expressions 
are provided in App.~\ref{app:potential}. To display the result,  
the dependence of the effective potential 
on the total energy $E_{\rm tot}^{(n)}$ is 
eliminated through initial-condition matching. 
For the purpose of illustration 
we choose the initial conditions $\theta(0)=0.05$, $\dot\theta(0)=-0.03\mathrm{s}^{-1}$, and  $\lambda=1/4$, $\rho=2$. 
Fig.~\ref{fig:Veff3} 
shows the effective potentials $V_{\rm eff}^{\rm LO}$, 
$V_{\rm eff}^{\rm NLO}$, $V_{\rm eff}^{\rm N3LO}$ obtained 
including the correction to $\ord{\lambda^2}$, $\ord{\lambda^4}$ and $\ord{\lambda^8}$, respectively, drawn in gray, blue and red, respectively. We see that the leading-power result deviates noticeably from the higher-order ones, which are already close to each other, indicating good convergence of the expansion for the chosen parameters.

\begin{figure}
 \centering
 \includegraphics[width=.45\textwidth]{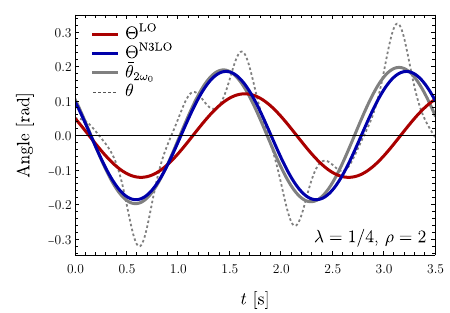}
 \caption{Motion of the pendulum from the subleading-power analysis, using the same parameters as for Fig.~\ref{fig:Veff3}. The dashed curve shows the exact numerical solution of the full system for comparison.}
 \label{fig:traj_slp}
\end{figure}

\begin{figure}
 \centering
 \includegraphics[width=.45\textwidth]{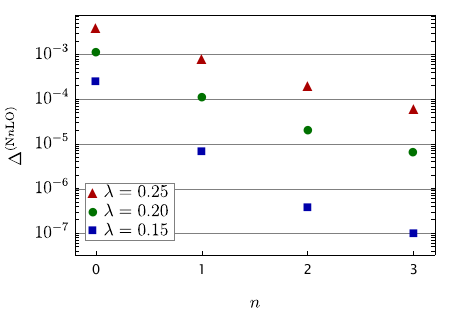}
 \caption{Deviations $\Delta^{(\mathrm{N}n\mathrm{LO})}$ of the effective and full solutions shown in Fig.~\ref{fig:traj_slp}, as defined in eq.~\eqref{eq:deviations}. For comparison, points for other values of $\lambda$ are also included.}
 \label{fig:deviations}
\end{figure}

From the effective potential we can next derive the effective motion of the pendulum  and compare it against the one obtained from the solution of eq.~\eqref{eq:eom_full}, which still contains the rapid oscillations. To do so, the initial conditions must be carefully adjusted order by order in $\lambda$ as detailed in Sec.~\ref{sec:initconds}. We show the averaged motions in Fig.~\ref{fig:traj_slp} at leading order, $\Theta^{\rm LO}(t)$, in red and $\Theta^{\rm N3LO}(t)$, obtained from the $\mathcal{O}(\lambda^8)$-accurate potential $V_{\rm eff}^{\rm N3LO}$, in blue. For comparison, we show the full solution $\theta(t)$ and its time-average $\bar\theta_{2\omega_0}(t)$, defined as $\theta(t)$ after applying a low-pass filter with cut-off frequency $2 \omega_0$ as the gray dashed and solid lines, respectively. In order to quantify the deviation of the effective solution from the exact one, we also determine the averaged square difference,
\begin{align}
 \Delta^{(\mathrm{N}n\mathrm{LO})} = \frac{1}{T}\int_0^T dt\, (\bar\theta_{2\omega_0}(t) - \Theta^{(\mathrm{N}n\mathrm{LO})}(t))^2\,, \label{eq:deviations}
\end{align}
and show the result in Fig.~\ref{fig:deviations}, also including different values for the expansion parameter $\lambda$, but keeping $\rho$ and the initial condition values unchanged. Rapid improvement with increasing order of approximation is seen in all three cases.

\subsection{Stability of the upright equilibrium}

\noindent Stability of the upright equilibrium requires 
\begin{align}
 \left. \frac{d^2}{d\Theta^2}V_\eff(\Theta,E_{\rm tot})\right|_{\Theta = 0} > 0\,. 
\label{eq:stability_general}
\end{align}
From the above expressions for $V_\eff^{(2)}$, 
$V_\eff^{(4)}$, we obtain 
\begin{align}
 \rho^2 > 2 + \frac{7}{4}\lambda^2 - \frac{12}{\omega^2}E_{\rm tot}^{(2)}
 +\ord{\lambda^4} \,,
 \label{eq:stability_NLP}
\end{align}
which includes the $\mathcal{O}(\lambda^2)$ correction 
to $\rho^2>2$. Interestingly, the inverted pendulum 
becomes stable already for smaller amplitude $h$ of the 
pivot point oscillation when the energy of the 
system increases. Yet, since the energy dependence is a subleading-power effect in $\lambda$, the depth of the potential well does not increase significantly. 

\begin{figure}
 \centering
 \includegraphics[width=.45\textwidth]{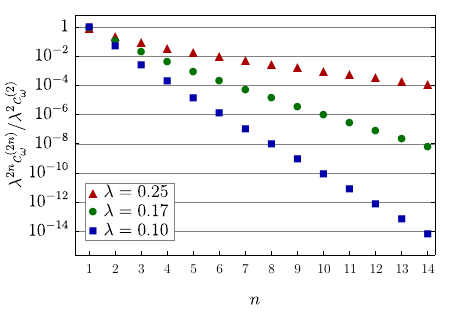}
 \caption{Relative corrections to the effective frequency as defined in eq.~\eqref{eq:effFreqSeries} with $\rho = 2$.}
 \label{fig:frequencies}
\end{figure}

If we are content with studying the frequency of a very small oscillation around the upright equilibrium point, we can obtain results to higher orders in $\lambda$.\footnote{The equation of motion of the full system in this limit takes the form of the well-known Mathieu equation.} For simplicity, we set $E_{\rm tot} = 0$ in the effective potential. In the limit of small amplitudes, the effective potential then directly yields the frequency of the (now sinusoidal) oscillation around the upright equilibrium, using
\begin{align}
 V_\eff = \frac{\omega_\eff^2 \Theta^2}{2} + \mathcal{O}(\Theta^3)\,.
\end{align}
From this we obtain the effective frequency $\omega_\eff$ of small-angle oscillations, which is expanded into a power-series in $\lambda$ as 
\begin{align}
 \omega_\eff^2  = \omega^2 \sum_{n=1}^\infty \lambda^{2n} c_\omega^{(2n)}\,. \label{eq:effFreqSeries}
\end{align}
Using computer algebra software, we calculated the 
expansion of $\omega_\eff^2$ up to order 
$\ord{\lambda^{28}}$, which corresponds to  
thirteen additional terms of the series beyond the leading approximation.
The first three coefficients are found to be
\begin{align}
\begin{aligned}
 c_\omega^{(2)} &= \frac{\rho^2}{2}-1 \,,\\
 c_\omega^{(4)} &= \frac{25}{32}\rho^4-2\rho^2 \,,\\
 c_\omega^{(6)} &= \frac{1169}{576}\rho^6 - \frac{273}{32}\rho^4 + 8\rho^2 \,.
\end{aligned} \label{eq:freqexp}
\end{align}
The complete set of calculated coefficients is 
given in App.~\ref {app:freq}.
Their magnitude relative to the leading term is shown in Fig.~\ref{fig:frequencies}. The individual terms are seen to fall off roughly exponentially with growing~$n$. 

\section{Convergence Behavior}

\noindent Equipped with a method to compute an effective Lagrangian to (in principle) any order in $\lambda$, it is interesting to investigate whether or not the series expansions converge. 
We consider the small oscillation frequency 
$\omega_{\rm eff}$, for which the first 14 terms of 
the series are at our disposal. In quantum mechanics 
(see \cite{BenderWu} for the anharmonic oscillator) and 
quantum field theory \cite{Dyson:1952tj}, perturbation 
expansions for energy eigenvalues or correlation 
functions usually have zero radius of convergence 
and the series are at best asymptotic. Inspection of 
the series coefficients for the classical mechanics 
problem at hand, however, suggests that there is a 
finite radius of convergence in $\lambda$. Failure 
of an expansion to converge often indicates that a qualitatively 
new physical phenomenon emerges at the convergence boundary. Indeed, one must expect 
the $\lambda$ expansion to fail for too large $\lambda=
\omega_0/\omega$ or $\lambda\rho=h/l$, and the effective description to break down, as the inverted 
pendulum exhibits parametric resonance \cite{Butikov_2011,ButikovStability2018,MOND199377,blackburn1992stability,insperger2002stability,insperger2003stability}, in which case 
the motion of the pendulum can certainly not be described 
by a time-independent effective Lagrangian with a conserved 
energy. 
This happens when the natural frequency $\omega_0$ 
of the pendulum and the driving frequency $\omega$ are 
close to each other, in which case there is no 
frequency separation and energy from the driver is transferred to the pendulum until it escapes the potential well around the upright equilibrium. 

To establish a baseline to which to compare the results from the effective description to high order in $\lambda$, we first investigate the stability of the full system by conducting a parameter scan of the trajectories in the $(\lambda\rho,\rho)$-plane. To simplify the following discussion and to make it more comparable to existing analyses on the Mathieu equation, we will focus solely on the small angle limit from now on. 
At each point, the full system is solved numerically 
with initial values $\theta(0)=0$ and $\dot{\theta}(0)=10^{-5}/\mathrm{s}$, and the time until the oscillation has amplified such 
that the pendulum tips over from the upright position (meaning it passes the point $|\theta|>\pi/2$) is measured. We solve the system until $t_\mathrm{max}=300\mathrm{s}$ and take $\omega$ for each parameter point such that $\omega_0 = \sqrt{g/l} = 3.14/\mathrm{s}$ (recall $\omega = \omega_0/\lambda$). If it does not tip over within the time scale of simulation, the point is considered stable. Choosing $t_{\rm max}=300\mathrm{s}$ turned out to be a reasonable criterion, because the typical tip-over time was orders of magnitude smaller than $t_{\rm max}$, apart from a region narrower than the provided resolution. 

\begin{figure}
\hskip-0.8cm
\includegraphics[width=0.5\textwidth]{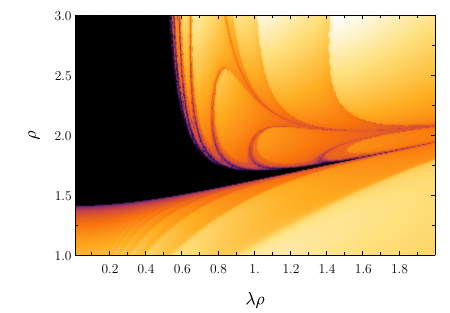}
    \caption{Stability of the numerical solutions in the $(\lambda\rho ,\rho)$ plane. Brighter colors denote solutions that tip over more quickly. The black region shows the parameter range where solutions are considered stable.}
    \label{fig:stability}
\end{figure}

We show the results of this simulation in Fig.~\ref{fig:stability}, where the coloring indicates the time at which the system tips over. Brighter colors denote solutions that tip over more quickly. The black wedge-shaped region is the parameter space in which the system is considered stable by the means of our simulation. The tip-over time exhibits interesting 
patterns in the instability region, which, however, are 
not our concern here. Instead, in what follows, we show how the effective description can be used to determine the boundary between the stable and unstable regime.

\begin{figure}
\hskip-0.8cm
\includegraphics[width=0.5\textwidth]{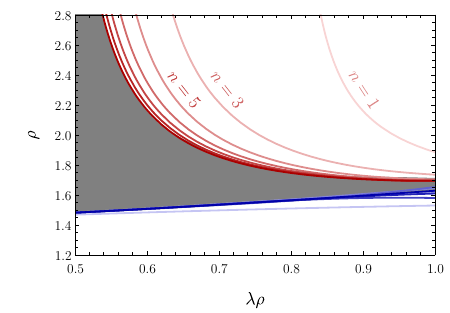}
    \caption{Boundary curves obtained from the stability condition $\omega_\eff^2>0$ (blue curves) and the convergence condition \eqref{eq:convergencecriterium} (red curves) with increasing accuracy in the $\lambda$ expansion. The shading of the different curves indicates the value of $n$ used in eq.~\eqref{eq:convergencecriterium}, with the most saturated color corresponding to the highest $n$. The dark gray area is the stable region.}
    \label{fig:stability_curves}
\end{figure}

Returning to the effective description, 
we consider the effective (small-angle) frequency $\omega_\eff$, defined in eq.~\eqref{eq:effFreqSeries}, 
setting $E_{\rm tot}=0$ in accordance with the initial 
conditions used to produce Fig.~\ref{fig:stability}. 
The lower boundary on $\rho$ for given $\lambda\rho$ corresponds to the stability condition 
which was provided by Kapitza at leading power ($\rho^2>2$). The inverted pendulum becomes unstable below that boundary 
because the driving amplitude is not sufficient to uphold the pendulum in the inverted position.  Solving eq.~\eqref{eq:effFreqSeries} for $\omega_\eff^2=0$ yields a condition on $\rho$ corresponding to 
the generalization of eq.~\eqref{eq:stability_NLP} to 
$\mathcal{O}(\lambda^{26})$. 
Once these higher-order corrections are included, the resulting curves approximate the lower boundary between the stable and unstable region found from the numerical solution to an excellent accuracy, as illustrated by the blue family of 
lines in Fig.~\ref{fig:stability_curves}. 

The upper boundary of the stability region in 
Fig.~\ref{fig:stability} coincides with the value of $\lambda\rho$ (given $\rho$), for 
which the $\lambda$ expansion of $\omega_\eff^2$ ceases 
to converge. We can estimate 
this value by requiring that the 
highest available term in the series \eqref{eq:effFreqSeries} is smaller than the previous 
one, which implies 
\begin{equation} 
\lambda < \sqrt{c_\omega^{(2n)}/c_\omega^{(2n+2)}}\,,
\label{eq:convergencecriterium}
\end{equation}
and quantifies the boundary 
of the region, where $\lambda$ can no longer be 
considered small enough to justify the effective treatment.
For illustration, we only show the algebraic result at next-to-leading order ($n=1$),
\begin{align}
 \lambda < \frac{4\sqrt{\rho^2-2}}{\rho\sqrt{25\rho^2-64}} \,,
\label{eq:convergence_NLP}
\end{align}
and work with numerical solutions at higher orders. Once these higher-order results are included, the resulting curves approximate the upper boundary between the stable and unstable region found from the numerical solution to an excellent accuracy 
with increasing approximation order $n$, as illustrated by the red family of lines in Fig.~\ref{fig:stability_curves}. 
The least saturated curve refers to $n=1$ and $n=13$ to the most saturated one. Since the effective Lagrangian approach 
assumes $\lambda, \lambda\rho \ll 1$, one expects that 
the boundary lines obtained from 
$\omega_{\rm eff}^2=0$ (blue) and 
\eqref{eq:convergencecriterium} (red) become unreliable 
for too large $\lambda\rho$, hence we do not plot them for 
$\lambda\rho >1$.

We can therefore conclude that the upright-equilibrium 
stability boundaries can be reliably inferred from the condition $\omega_{\rm eff}^2=0$  
and from the convergence of the series expansion of  $\omega_{\rm eff}^2$ in the effective potential method.

The above findings are an interesting way to study the stability of the Mathieu functions (i.e. the solutions to the linearized equation of motion of the system) using effective theories. Another approach which has  been adopted in the literature~\cite{blackburn1992stability,MOND199377,insperger2002stability,insperger2003stability,Butikov_2011,ButikovStability2018,kovacic2018mathieu}, relates the existence of stable, 
oscillatory solutions to the frequency separation. 
In our framework this approach can be stated as follows: When we assume that the mode functions $\Phi_n$ and $\Psi_n$ vary with time only over the time scale of $\omega_\eff$, such that the derivatives scale as powers of $\lambda$, we implicitly make the assumption that $\omega_\eff \sim \lambda\omega=\omega_0$. However, as can be 
seen from \eqref{eq:effFreqSeries}, this is no longer 
the case when $\rho$ becomes increasingly large, which 
corresponds to the situation when the amplitude $h$ of the 
pivot point oscillation is no longer of order $\lambda$ 
relative to the length of the pendulum. From the physical 
point of view, the criterion for mode separation is 
not $\omega\gg \omega_0$ but $\omega\gg \omega_\eff$. 
When $\lambda\rho$ is large enough such that 
\begin{equation}
\omega_1^-\equiv \omega-\omega_\eff\sim \omega_\eff\,,
\end{equation}
the effective treatment based on fast and slow frequency 
separation is certainly no longer 
applicable. 
Equating $\omega_1^{-}=\omega_\eff$ and inserting the leading-order expression for $\omega_\eff=\lambda\omega \sqrt{\rho^2/2-1}$ as an estimate, yields an approximation for when this occurs, namely at
\begin{align}
 \lambda_{\rm crit} \gtrsim  \frac{1}{\sqrt{2}}(\rho^2-2)^{-1/2}  \,,
\end{align}
which at $\rho=2$ amounts to $\lambda_\mathrm{crit}=1/2$.

This method provides yet another way of identifying parameter regions in which the effective description cannot converge. Using our $\ord{\lambda^{28}}$-expression for $\omega_\eff$, and solving the relation $\omega_1^{-}=\omega_\eff$ for $\lambda$ numerically for a given $\rho$, we find that the resulting curves 
are virtually indistinguishable from the red lines in Fig.~\ref{fig:stability_curves} (and therefore not drawn), although their algebraic 
expressions are not identical.


\section{Conclusion}

\noindent In this paper, we applied the effective Lagrangian 
method beyond the leading approximation to a classical problem, the driven inverted pendulum 
with driving frequency $\omega$ much higher than the one of the unperturbed pendulum, $\omega_0$. Using methods inspired by scale separation and power-counting in effective quantum field theories, we derived a general algorithm, valid to any order in the scale separation parameter $\lambda = \omega_0/\omega$, in which the systematic expansion is constructed. The approach employs a modulated Fourier mode expansion and generalizes the procedure found in textbooks \cite{LandauLifshitz}, which is based on averaging high frequency modes.

The effective Lagrangian obtained at higher orders in $\lambda$ is time-independent, but exhibits interesting properties that would be considered unconventional and even problematic for a general mechanical system, as it depends on high powers of the derivative of the generalized coordinate, as well as on its higher derivatives. These two properties translate to the system not having a potential of the form $V(\Theta)$, as well as possibly having an unbounded Hamiltonian. However, the problematic terms appear only in the Lagrangian at higher order in the expansion parameter. This allows us to treat them as perturbative corrections and in turn eliminate them in such a way that the full system is canonical and has an effective potential $V_{\rm eff}(\Theta,E_{\rm tot})$. 
The effective potential depends however on the total energy of the system. By energy conservation following from the time-independence of the effective Lagrangian, this feature can be rephrased as a dependence on the initial conditions. The initial 
conditions themselves need to be carefully matched between 
the effective system with generalized coordinate $\Theta$ 
and the full system with coordinate $\theta$.

We computed the effective potential explicitly including 
the first three corrections to the previously known 
leading expression, and the curvature of the potential 
in the upright equilibrium position to order 
$\lambda^{28}$, corresponding to the first 14 terms
terms of the series.

By studying the effective potential at such high orders in the expansion parameters, we estimated the parameter regions of stable trajectories to high accuracy. This method allows us to determine the parameter-space region in which parametric resonance occurs, the effective Lagrangian method breaks down, and the upright equilibrium is unstable. We find that this approach yields comparable results to the one found in previous studies from imposing sufficient separation of frequencies. \\[0.3cm]

\section*{Acknowledgments}
\noindent M.B. thanks A.~Penin for making him aware of the inverted 
pendulum many years ago. The software FORM~\cite{vermaseren2000new} was used for the high-order calculations. 
This work has been supported in part by the Excellence 
Cluster 
ORIGINS funded by the Deutsche Forschungsgemeinschaft 
(DFG, German Research Foundation) under Ger\-many's Excellence Strategy --EXC-2094 --390783311. 
%
%
\appendix
\onecolumngrid

\section{Higher-Order Effective Potential}
\label{app:potential}

\noindent We give the subleading expressions for the effective potentials $V_\mathrm{eff}^{(n)}$. For notational convenience, we introduce the rescaled energy expansion coefficients $k_n = E^{(n)}_\mathrm{tot}/\left(\lambda^{n} \omega^2\right)$, where $E^{(n)}_\mathrm{tot}$ again refers to the $\lambda^n$ contribution of the total energy. 
We obtain the following potentials up to the third subleading correction:
\small
{\allowdisplaybreaks
\begin{align*}
		\frac{V_\mathrm{eff}^{(2)}}{\lambda^2 \omega^2}\	=&  \ c_{\Theta }-1 +\rho ^2 \left(-\frac{1}{8} c_{2 \Theta }+\frac{1}{8}\right)
		\\
		\frac{V_\mathrm{eff}^{(4)}}{\lambda^4 \omega^2}\	=& \  \rho ^2 \left(\frac{17}{16} c_{\Theta }-\frac{3}{4} c_{2 \Theta }+\frac{7}{16} c_{3 \Theta }-\frac{3}{4}+k_2 \left[-\frac{3}{4} c_{2 \Theta }+\frac{3}{4}\right]\right) \\ &+\rho ^4 \left(\frac{1}{32} c_{2 \Theta }-\frac{29}{512} c_{4 \Theta }+\frac{13}{512}\right)
		\\
		\frac{V_\mathrm{eff}^{(6)}}{\lambda^6 \omega^2}\   =& \ \rho ^2 \left(\frac{87}{4} c_{\Theta }-\frac{25}{2} c_{2 \Theta }+\frac{33}{4} c_{3 \Theta }-\frac{27}{8} c_{4 \Theta }-\frac{113}{8}\right.\\
		& \qquad \quad +\left.k_2 \left[\frac{87}{4} c_{\Theta }-5 c_{2 \Theta }+\frac{33}{4} c_{3 \Theta }-25\right]+k_2^2 \left[-\frac{5}{2} c_{2 \Theta }+\frac{5}{2}\right]+k_4 \left[-\frac{3}{4} c_{2 \Theta }+\frac{3}{4}\right]\right)\\ 
		&+\rho ^4 \left(\frac{2167}{2048} c_{\Theta }-\frac{123}{32} c_{2 \Theta }+\frac{10665}{4096} c_{3 \Theta }-\frac{629}{256} c_{4 \Theta }+\frac{6345}{4096} c_{5 \Theta }+\frac{279}{256}\right.\\
		&\qquad \quad+ \left.k_2 \left[-\frac{87}{32} c_{2 \Theta }-\frac{629}{256} c_{4 \Theta }+\frac{1325}{256}\right]\right) \\
		&+\rho ^6 \left( \frac{20183}{73728} c_{2 \Theta }+\frac{243}{2048} c_{4 \Theta }-\frac{1143}{8192} c_{6 \Theta }-\frac{4661}{18432}\right) \\
		\frac{V_\mathrm{eff}^{(8)}}{\lambda^8\omega^2}\ = & \ \rho ^2 \left( \frac{1981}{4} c_{\Theta }-371 c_{2 \Theta }+\frac{1989}{8} c_{3 \Theta }-\frac{605}{4} c_{4 \Theta }+\frac{385}{8} c_{5 \Theta } -\frac{1079}{4} \right.\\
		& \qquad \quad +\left.k_2 \left[\frac{1211}{2} c_{\Theta }-385 c_{2 \Theta }+\frac{429}{2} c_{3 \Theta }-\frac{605}{4} c_{4 \Theta }-\frac{1135}{4}\right] \right.\\
		& \qquad \quad +\left.k_2^2 \left[\frac{1211}{4} c_{\Theta }-21 c_{2 \Theta }+\frac{429}{4} c_{3 \Theta }-389\right] +k_2^3 \left[-7 c_{2 \Theta }+7\right]\right.\\
		& \qquad \quad +\left. k_4 \left[\frac{87}{4} c_{\Theta }-5 c_{2 \Theta }+\frac{33}{4} c_{3 \Theta }-25\right]+k_2 k_4 \left[-5 c_{2 \Theta }+5\right]+k_6 \left[-\frac{3}{4} c_{2 \Theta }+\frac{3}{4}\right]\right)\\
		&+\rho ^4 \left(\frac{15903}{128} c_{\Theta }-\frac{7836379}{32768} c_{2 \Theta }+\frac{499467}{2048} c_{3 \Theta }-\frac{2977261}{16384} c_{4 \Theta } \right. \\
		& \qquad \quad +\left.\frac{261525}{2048} c_{5 \Theta }-\frac{1557909}{32768} c_{6 \Theta }-\frac{449115}{16384}\right.\\
		& \qquad \quad +\left.k_2 \left[\frac{11727}{128} c_{\Theta }-283 c_{2 \Theta }+\frac{474123}{2048} c_{3 \Theta }-\frac{4627}{32} c_{4 \Theta }+\frac{261525}{2048} c_{5 \Theta }-\frac{2973}{128}\right] \right.\\
		& \qquad \quad +\left.k_2^2 \left[-\frac{253}{2} c_{2 \Theta }-\frac{4627}{64} c_{4 \Theta }+\frac{12723}{64}\right] +k_4 \left[-\frac{87}{32} c_{2 \Theta }-\frac{629}{256} c_{4 \Theta }+\frac{1325}{256}\right]\right) \\
		&+\rho ^6 \left(\frac{44553629}{5308416} c_{\Theta }+\frac{2503511}{110592} c_{2 \Theta }-\frac{39310133}{5308416} c_{3 \Theta }-\frac{81293}{9216} c_{4 \Theta }\right.\\
		&\qquad \quad +\left.\frac{1665323}{196608} c_{5 \Theta } -\frac{71277}{4096} c_{6 \Theta }+\frac{749151}{65536} c_{7 \Theta }-\frac{478423}{27648} \right. \\
		&\qquad \quad +\left.k_2 \left[\frac{3386735}{110592} c_{2 \Theta }-\frac{47327}{9216} c_{4 \Theta }-\frac{71277}{4096} c_{6 \Theta }-\frac{223583}{27648}\right]\right) \\
		&+ \rho ^8 \left(-\frac{7968835}{2654208} c_{2 \Theta }+\frac{9568931}{10616832} c_{4 \Theta }+\frac{333907}{294912} c_{6 \Theta }-\frac{1622357}{2097152} c_{8 \Theta }+\frac{295983029}{169869312}\right)
\end{align*}
}

\section{Higher-Order Initial Conditions}
\label{app:init_conds}

\noindent We provide the matched initial condition to the third subleading power. For simplicity we set the initial time $t_i=0$ and define $ \theta_i\equiv \theta(0)$. We also define the rescaled initial velocity of the full system $\nu_i\equiv \frac{\dot{\theta}(0)}{\lambda \omega }$. The expansion of the initial conditions for the generalized coordinate $\Theta(t)$ of the effective description takes the form
\begin{equation*}
	\begin{aligned}
		\Theta(0) &= \theta_i + \sum_{n=1}\Theta_i^{(2n)}\,, \\
		\dot{\Theta}(0) &= \dot{\theta}_i - \lambda\rho \omega s_{\theta_i} + \sum_{n=1}\dot{\Theta}_i^{(2n+1)}\,,
	\end{aligned}
\end{equation*}
where we defined the $\mathcal{O}(\lambda^{2n})$ contribution to the effective initial condition $\Theta(0)$ as $\Theta_i^{(2n)}$, similarly for its derivative. In the following expressions for the expansion coefficients, we use the abbreviations $\nu_i\equiv \frac{\dot{\theta}(0)}{\lambda \omega }$ and $\mu_i^{(2n)} \equiv \dot{\Theta}_i^{(2n+1)}/(\lambda\omega)$:
\small
{\allowdisplaybreaks
\begin{align*}
		\frac{\Theta_i^{(2)}}{\lambda^2} \ =& \ -2 \rho c_{\theta _i} \nu _i + \frac{17}{16} \rho ^2  s_{2 \theta _i}
		\\
		\frac{\mu_i^{(2)}}{\lambda^2}\ =&  \ \rho \biggl(s_{2 \theta _i}-\nu _i^2 s_{\theta _i}\biggr)+\rho ^2 \left(\nu _i \left[-\frac{3}{8} c_{2 \theta _i}+\frac{7}{4}\right]\right)+\rho ^3 \left(-\frac{53}{48} s_{\theta _i}-\frac{5}{16} s_{3 \theta _i}\right)
		\\
		\frac{\Theta_i^{(4)}}{\lambda^4}\ =&  \ \rho \biggl(\nu _i \left[11 c_{2 \theta _i}-3\right]-4 \nu _i^3 c_{\theta _i}\biggr)+\rho ^2 \left(\frac{947}{128} s_{\theta _i}-\frac{865}{128} s_{3 \theta _i}+\frac{45}{8} \nu _i^2 s_{2 \theta _i}\right) \\
		&+\rho ^3 \left(\nu _i \left[-\frac{409}{108} c_{\theta _i}-\frac{3}{2} c_{3 \theta _i}\right]\right)+\rho ^4 \left(-\frac{1529}{3456} s_{2 \theta _i}+\frac{4039}{2048} s_{4 \theta _i}\right)
		\\
		\frac{\mu_i^{(4)}}{\lambda^4}\ =& \ \rho \biggl(s_{\theta _i}-3 s_{3 \theta _i}+11 \nu _i^2 s_{2 \theta _i}-\nu _i^4 s_{\theta _i}\biggr) \\
		&+\rho ^2 \left(\nu _i \left[-\frac{2067}{128} c_{\theta _i}+\frac{899}{128} c_{3 \theta _i}\right]+\nu _i^3 \left(-\frac{5}{4} c_{2 \theta _i}+\frac{11}{2}\right)\right) \\ &+\rho ^3 \left(\frac{21961}{3456} s_{2 \theta _i}+\frac{1319}{768} s_{4 \theta _i}+\nu _i^2 \left[-\frac{5867}{432} s_{\theta _i}-\frac{75}{16} s_{3 \theta _i}\right]\right) \\
		&+\rho ^4 \left(\nu _i \left[\frac{3361}{1728} c_{2 \theta _i}-\frac{1785}{512} c_{4 \theta _i}+\frac{42833}{6912}\right]\right) \\
		&+\rho ^5 \left(-\frac{10507}{34560} s_{\theta _i}-\frac{66475}{27648} s_{3 \theta _i}+\frac{139}{3072} s_{5 \theta _i}\right)
		\\
		\frac{\Theta_i^{(6)}}{\lambda^6}\ =& \ \rho \biggl(\nu _i \left[58 c_{\theta _i}-90 c_{3 \theta _i}\right]+\nu _i^3 \left[114 c_{2 \theta _i}-10\right]-6 \nu _i^5 c_{\theta _i}\biggr) \\
		&+\rho ^2 \left(-\frac{19437}{256} s_{2 \theta _i}+\frac{28125}{512} s_{4 \theta _i}+\nu _i^2 \left[\frac{100257}{512} s_{\theta _i}-\frac{86739}{512} s_{3 \theta _i}\right]+\frac{259}{16} \nu _i^4 s_{2 \theta _i}\right) \\
		&+\rho ^3 \left(\nu _i \left[\frac{5098439}{31104} c_{2 \theta _i}-\frac{33565}{13824} c_{4 \theta _i}-\frac{4253917}{41472}\right] \right. \\
		& \qquad \quad \left.+\nu _i^3 \left[-\frac{13963}{486} c_{\theta _i}-\frac{359}{6} c_{3 \theta _i}\right]\right) \\
		&+\rho ^4 \left(\frac{480145957}{7962624} s_{\theta _i}+\frac{8066321}{1990656} s_{3 \theta _i}-\frac{29580613}{884736} s_{5 \theta _i}\right. \\
		&\qquad \quad \left.+\nu _i^2 \left[-\frac{212263}{2592} s_{2 \theta _i}+\frac{53107}{512} s_{4 \theta _i}\right]\right) \\
		&+\rho ^5 \left(\nu _i \left[\frac{95375231}{1944000} c_{\theta _i}-\frac{2177353}{20736} c_{3 \theta _i}+\frac{125921}{3456} c_{5 \theta _i}\right]\right) \\
		&+\rho ^6 \left(-\frac{39063010973}{1327104000} s_{2 \theta _i}+\frac{10440595}{497664} s_{4 \theta _i}+\frac{1506521}{3538944} s_{6 \theta _i}\right)
		\\
		\frac{\mu_i^{(6)}}{\lambda^6}\ =& \ \rho\biggl(-18 s_{2 \theta _i}+17 s_{4 \theta _i}+\nu _i^2 \left[29 s_{\theta _i}-135 s_{3 \theta _i}\right]+57 \nu _i^4 s_{2 \theta _i}-\nu _i^6 s_{\theta _i}\biggr) \\
		&+\rho ^2 \left(\nu _i \left[\frac{23917}{128} c_{2 \theta _i}-\frac{13501}{128} c_{4 \theta _i}-\frac{117}{4}\right] \right.\\
		&\qquad \quad +\left.\nu _i^3 \left[-\frac{103179}{512} c_{\theta _i}+\frac{41811}{512} c_{3 \theta _i}\right]+\nu _i^5 \left[-\frac{21}{8} c_{2 \theta _i}+\frac{45}{4}\right]\right) \\
		&+\rho ^3 \left(\frac{4069973}{62208} s_{\theta _i}-\frac{1231805}{20736} s_{3 \theta _i}-\frac{2369}{432} s_{5 \theta _i}\right.\\
		&\qquad \quad +\left. \nu _i^2 \left[\frac{29980835}{124416} s_{2 \theta _i}+\frac{2973521}{27648} s_{4 \theta _i}\right]+\nu _i^4 \left[-\frac{113107}{1944} s_{\theta _i}-\frac{419}{8} s_{3 \theta _i}\right] \right) \\
		&+\rho ^4 \left(\nu _i \left[-\frac{725855525}{7962624} c_{\theta _i}-\frac{174397183}{1990656} c_{3 \theta _i}+\frac{98781193}{884736} c_{5 \theta _i}\right]\right.\\
		&\qquad \quad +\left. \nu _i^3 \left[\frac{418565}{3888} c_{2 \theta _i}-\frac{10511}{128} c_{4 \theta _i}+\frac{1025965}{15552}\right]\right) \\
		&+\rho ^5 \left(-\frac{60743917147}{1990656000} s_{2 \theta _i}+\frac{5150996737}{79626240} s_{4 \theta _i}-\frac{107574157}{8847360} s_{6 \theta _i}\right.\\
		&\qquad \quad +\left.\nu _i^2 \left[\frac{485775337}{7776000} s_{\theta _i}-\frac{15323549}{82944} s_{3 \theta _i}+\frac{283495}{27648} s_{5 \theta _i}\right]\right) \\
		&+\rho ^6 \left(\nu _i \left[\frac{233925934847}{1990656000} c_{2 \theta _i}-\frac{19863253}{497664} c_{4 \theta _i}-\frac{32679059}{1769472} c_{6 \theta _i}-\frac{1259207923}{31104000}\right]\right) \\
		&+\rho ^7 \left(\frac{183455603621}{5573836800} s_{\theta _i}-\frac{67196055007}{3981312000} s_{3 \theta _i}-\frac{234812363}{31850496} s_{5 \theta _i}+\frac{63757471}{17694720} s_{7 \theta _i}\right)
\end{align*}
}

\section{Harmonic Frequency Expressions}
\label{app:freq}

\noindent We give the expansion coefficients for the effective harmonic frequency $\omega_{\rm eff}^2$ defined in \eqref{eq:effFreqSeries}. We calculated the first 14 terms of the expansion, which corresponds to $\lambda^{28}$. The  coefficients are polynomials in $\rho$ with rational coefficients. 

\tiny
{\allowdisplaybreaks
	\begin{align*}
		c_\omega^{(2)} =& \frac{1}{2} \rho ^2-1
		\\
		c_\omega^{(4)} =& \frac{25}{32} \rho ^4-2 \rho ^2
		\\
		c_\omega^{(6)} =& \frac{1169}{576} \rho ^6-\frac{273}{32} \rho ^4+8 \rho ^2
		\\
		c_\omega^{(8)} =& \frac{16824665}{2654208} \rho^8-\frac{94831}{2592} \rho^6+\frac{2049}{32} \rho^4-32 \rho^2
		\\
		c_\omega^{(10)} =&  \frac{104550461873}{4777574400} \rho^{10}-\frac{15301534105}{95551488} \rho^8+\frac{18943067}{46656} \rho^6-\frac{13057}{32} \rho^4+128 \rho^2
		\\
		c_\omega^{(12)} =& \frac{1383860829361699}{17199267840000} \rho ^{12}-\frac{510599414365027}{716636160000} \rho ^{10}+\frac{2704847264795}{1146617856} \rho ^8-\frac{187198453}{52488} \rho ^6+\frac{75777}{32} \rho ^4-512 \rho ^2 
		\\
		c_\omega^{(14)} =& \frac{29289023958538918009}{94810963968000000} \rho ^{14}-\frac{12419023504555535053}{3869835264000000} \rho ^{12}+\frac{16897194776220398401}{1289945088000000} \rho ^{10}-\frac{3275894575046105}{123834728448} \rho ^8\\ &+\frac{102745082885}{3779136} \rho ^6 -\frac{413697}{32} \rho ^4+2048 \rho ^2
		\\
		c_\omega^{(16)} =& \frac{83857909128685125912504026393}{68504182964040499200000000} \rho ^{16}-\frac{488045135705765455513091071}{33449308087910400000000} \rho ^{14}+\frac{985874975154709612644967}{13931406950400000000} \rho ^{12} \\ &-\frac{155733295202139312167933}{870712934400000000} \rho ^{10} +\frac{1114567661360939425}{4458050224128} \rho ^8-\frac{3200742128665}{17006112} \rho ^6+\frac{2162689}{32} \rho ^4-8192 \rho ^2
		\\
		c_\omega^{(18)} =& \frac{120089176174908625267123147897106657}{24168275749713488117760000000000} \rho ^{18}-\frac{269168508120402223082153608059967051}{4028045958285581352960000000000} \rho ^{16} \\ &+\frac{92279593282170513811138202650217}{245852414446141440000000000} \rho ^{14} -\frac{3571425412964887484418187991}{3134566563840000000000} \rho ^{12}+\frac{2108768571265773123745227511}{1044855521280000000000} \rho ^{10} \\ &-\frac{112207057270371335395}{53496602689536} \rho ^8 +\frac{371789779789295}{306110016} \rho ^6-\frac{10944513}{32} \rho ^4+32768 \rho ^2
		\\
		c_\omega^{(20)} =& \frac{175263403904227897478594054729227794909043}{8526567684498918607945728000000000000} \rho ^{20}-\frac{131243335963394451160758386540543920310989}{426328384224945930397286400000000000} \rho ^{18} \\ &+\frac{279232015345030240850013025618323421714447}{142109461408315310132428800000000000} \rho ^{16}-\frac{45296196240036131201719335763017237187}{6505254886244902502400000000000} \rho ^{14} \\ &+\frac{33739061859510881881535523290359}{2256887925964800000000000} \rho ^{12}-\frac{9355356107413400058873181771417}{470184984576000000000000} \rho ^{10}+\frac{93058155781360081771505}{5777633090469888} \rho ^8 \\ &-\frac{5116060734204095}{688747536} \rho ^6+\frac{54001665}{32} \rho ^4-131072 \rho ^2
		\\
		c_\omega^{(22)} =& \frac{282847525631781705724885459394868767290384277872373}{3275900483130336930037862341017600000000000000} \rho ^{22}-\frac{19290040218535541559219833629556583270732139031503}{13536778855910483181974637772800000000000000} \rho ^{20} \\ &+\frac{13788286801022032955390760875602758321381259880239}{1353677885591048318197463777280000000000000} \rho ^{18}-\frac{3103818637959156164712319406632069993514674355453}{75204326977280462122081320960000000000000} \rho ^{16} \\ &+\frac{7470836385204179119168031963532059214635417}{71720435120850050088960000000000000} \rho ^{14}-\frac{430039973727914354734435397867644739}{2538998916710400000000000000} \rho ^{12} \\ &+\frac{449916002199871703719766958103238663}{2538998916710400000000000000} \rho ^{10}-\frac{24019279122807117265238905}{207994791256915968} \rho ^8 \\ &+\frac{1079536847309190425}{24794911296} \rho ^6-\frac{261095425}{32} \rho ^4+524288 \rho ^2
		\\
		c_\omega^{(24)} =& \frac{462268189407622980070187142794131930972283406622989689224253}{1258598344898288944251002681129089105920000000000000000} \rho ^{24} \\ &-\frac{86826106905618634905556618093828366949132731009071000532877}{13110399426023843169281277928428011520000000000000000} \rho ^{22} \\ &+\frac{93847416638175046132782637159577187302841621594668517389}{1790915842636956924975244577341440000000000000000} \rho ^{20} \\ &-\frac{128387128420533705367916342309325844669854222563018198989}{537274752791087077492573373202432000000000000000} \rho ^{18} \\ &+\frac{1022077886396251550445416469859905138490704725080596389}{1474004808754697057592793890816000000000000000} \rho ^{16} \\ &-\frac{6248781427949306780090443236918867305119283092831}{4685735094562203272478720000000000000000} \rho ^{14} +\frac{5229382962963418635872640393212715850861}{3046798700052480000000000000000} \rho ^{12} \\ &-\frac{46248045973513272796578928600044367951}{31737486458880000000000000000} \rho ^{10} +\frac{217332854394611010157695955}{277326388342554624} \rho ^8-\frac{27507355568072365315}{111577100832} \rho ^6 \\ &+\frac{1241513985}{32} \rho ^4-2097152 \rho ^2
		\\
		c_\omega^{(26)} =& \frac{129038512518653608908013841918133236441640681858146168536514742761447177}{81720368652080691243763491149613041176717295616000000000000000000} \rho ^{26} \\ &-\frac{7466672475271258206411032946255392638581353745701451188541219236031989}{241776238615623346875039914643825565611589632000000000000000000} \rho ^{24} \\ &+\frac{5398845519120437529448129129031926002295887567322877054887392159172379}{20148019884635278906253326220318797134299136000000000000000000} \rho ^{22} \\ &-\frac{23168943557410693360731764696486962814988326646666342872117918209}{17059547950622596884544189745923620864000000000000000000} \rho ^{20} \\ &+\frac{15199963329184432993009020771898106646677055890472950214498023017}{3411909590124519376908837949184724172800000000000000000} \rho ^{18} \\ &-\frac{12869476934636564747653863325226820557044411384935409167629513}{1300072241321642804796844211699712000000000000000000} \rho ^{16} \\ &+\frac{14089529769328588448840379000605568208323725234457151441}{929884129515869239423401984000000000000000000} \rho ^{14} \\ &-\frac{10915670389386472594378595790158606675747899}{685529707511808000000000000000000} \rho ^{12} +\frac{2567346427624746880813248711133145887270177}{228509902503936000000000000000000} \rho ^{10} \\ &-\frac{1370308410719375473886395601225}{269561249468963094528} \rho ^8+\frac{2724400629069976643459}{2008387814976} \rho ^6 -\frac{5821693953}{32} \rho ^4+8388608 \rho ^2
		\\
		c_\omega^{(28)} =& \frac{36345946021027445341629359633703104737621781477317320811023948708058852109306767193}{5306076143912027105009858570822141413312851568709389516800000000000000000000} \rho ^{28} \\ &-\frac{27369000728578716214966200477234554367256981455962616682322216498296622080436149283}{189502719425429539464637806100790764761173270311049625600000000000000000000} \rho ^{26} \\ & \\ &+\frac{9046641423542327760184622591380741775142835883666872154875740353786511874617599}{6635016961080828383622345369587576231965731953049600000000000000000000} \rho ^{24} \\ &-\frac{6306250910017840861488192420418119232469566746100719153263715392889596030343689}{829377120135103547952793171198447028995716494131200000000000000000000} \rho ^{22} \\ &+\frac{215301979372622936192888758236579557739971628689805217012783163354456239}{7738210950402409946829244468750954423910400000000000000000000} \rho ^{20} \\ &-\frac{135744408063298052316132454385830653034704388562887895544837073042367557}{1934552737600602486707311117187738605977600000000000000000000} \rho ^{18} \\ &+\frac{12250032304270105368254748800235591753615443417621184518084360837597}{98285461443916196042641422404498227200000000000000000000} \rho ^{16} \\ &-\frac{3209650102469216338239811956334222922347371427779307065627301}{20503945055824916729286013747200000000000000000000} \rho ^{14}\\&+\frac{339268309048445977552318981024838353042166016577}{2467906947042508800000000000000000000} \rho ^{12} -\frac{76004060694084228949294349897409420899453027821}{925465105140940800000000000000000000} \rho ^{10} \\ &+\frac{308135893627892285068404828937105}{9704204980882671403008} \rho ^8-\frac{8230657846197335693237}{1129718145924} \rho ^6+\frac{26977763329}{32} \rho ^4-33554432 \rho ^2
\end{align*}
}
\normalsize

\bibliography{bibliography}

\end{document}